\documentclass[12pt]{iopart}
\usepackage{graphicx}
\usepackage{rotating}
\usepackage{dcolumn}
\usepackage{epsfig}

\usepackage{amssymb}

\begin{document}

\title{Network Mutual Information and Synchronization under Time Transformations}

\author{T. Pereira$^{1}$\footnote{Present address: Physics Institute, University of S\~ao Paulo, Brazil, Rua do Mat\~ao Travessa R}, M. Thiel$^{2}$, M.S. Baptista$^3$,and J.  Kurths$^{1}$} 
  
\address{ $^1$ Nonlinear Dynamics Group, Institute of Physics University of
  Potsdam, D-14415, Potsdam, Germany} 

\address{$^2$ Centre for Applied Dynamics Research, King's College, Aberdeen, AB24 3UE, UK }
  
\address{ $^3$ Max-Planck Institute f\"ur Physik
 komplexer Systeme, N\"othnitzerstr. 38, D-01187 Dresden, Deutschland}

\ead{tiagops@if.usp.br}

\date{\today}

\begin{abstract}

We investigate the effect of general time transformations on the
phase synchronization phenomenon and the mutual information rate between pairs
of nodes in dynamical networks. We demonstrate two important results
concerning the invariance of both phase synchronization and the mutual
information rate. Under time transformations phase
synchronization can neither be introduced nor destroyed and the mutual
information rate cannot be raised from zero.  On the other hand, 
for proper time transformations the timing between the cycles of the coupled oscillators
can be largely improved. Finally, we discuss the
relevance of our findings for communication in dynamical networks.

\end{abstract}

\maketitle

\section{Introduction}

Time and complex dynamics play a major role in biological, social, economical,
and physical systems.  Cycles of different periods often govern their
dynamical behavior and determine their intrinsic activity.  A variety of
processes require a precise timing between the oscillators cycles for a proper
functioning, as for example, the respiratory and cardiac systems
\cite{Schaefer}, spike discharges and information transmission
\cite{Riehle,borisyuk} in neuron networks, ecology \cite{blasius}, fireflies
blinking after dark, and peacemaker cells of the human heart \cite{Strogatz}.
Synchronization is an efficient mechanism to generate such a timing
\cite{Schaefer,Riehle,borisyuk,blasius,Strogatz,fujisaka,CommSync,livro,Varela,Gruen,Lai}.  
Among several types of synchronization recently found in complex systems \cite{livro},
chaotic phase synchronization (PS) displays special importance because of its
weak constraints on the dynamics and coupling strength.  It has been reported
that PS mediates the process of information transmission and collective
behavior in neural and active networks \cite{Riehle,Lai,Murilo-Canal}, as well
as communication processes in the human brain
\cite{Riehle,parkinson,mormann:2003}.

In real systems PS is the most common type of synchronization
\cite{Schaefer,Riehle,blasius,Strogatz,livro,Varela,Gruen,Lai,parkinson,mormann:2003}. The
main reason for PS to be so common relays on the fact that real
oscillators are not identical, but have some parameter mismatch. When
real coupled oscillators undergo a transition to PS the timing is not
precise. In many situations one wishes to improve the timing
condition, but in fact, one cannot systematically control the
oscillator parameters to drive them to a higher level of PS. The
question is then how to improve the timing without changing the
oscillator parameters. The natural candidate is a time
transformation. Could one enhance a better timing by changing the
time? Or even better, could one introduce PS by time
transformations?

Coupled dynamical systems under time transformations are important
in physics without an absolute time as well as in situations where the time 
cannot be directly obtained, as in the study of sedimental cores in the field of
geophysics.  In the latter case, the time at which the sedimentation took
place is usually unknown.  Only a proxy for the time can be derived from the
measurements, which does not yield the "real" time but only a monotonous
transformation of it \cite{marco}.  In the study of synchronization phenomenon
in such a system the natural question is whether not having access to the {\it
real} time could effect synchronization.

Such time transformations (typically nonlinear) have attracted a great deal of
attention (see  \cite{Adilson,fayad} and references therein). They cause no change in the topology of the dynamics, 
but the duration of the cycles can be drastically modified.
 An important problem is to analyze whether the dynamical properties
are invariant under time transformations \cite{Adilson}.  Recent results have
shown that dynamical systems under time transformation can present
nontrivial and counterintuitive properties. For example, a nonmixing dynamics
can be converted to a mixing one \cite{fayad}.

In this work, we show that time transformations, satisfying simple
conditions of integrability, can neither introduce nor destroy the phenomenon
of PS. We also explore the natural connection between
synchronization and information exchange in coupled oscillators. We uncover
the transformation law for the mutual information rate (MIR), the rate with
which information about a node can be retrieved in another node. If
the MIR is zero in one time frame it will remain zero for any other. 
On the other hand, if the MIR is nonzero it can be drastically modified by a 
time transformation . {\it Surprisingly, if there is no synchronization (to any extent) 
between the nodes forming a network, time transformations containing 
information about a particular oscillator (node) of the network cannot be used to carry this information 
to another oscillator}.

\section{Two Oscillators Case: Enhancing a Precise Timing}

\begin{figure}[h]
  \centerline{\hbox{\psfig{file=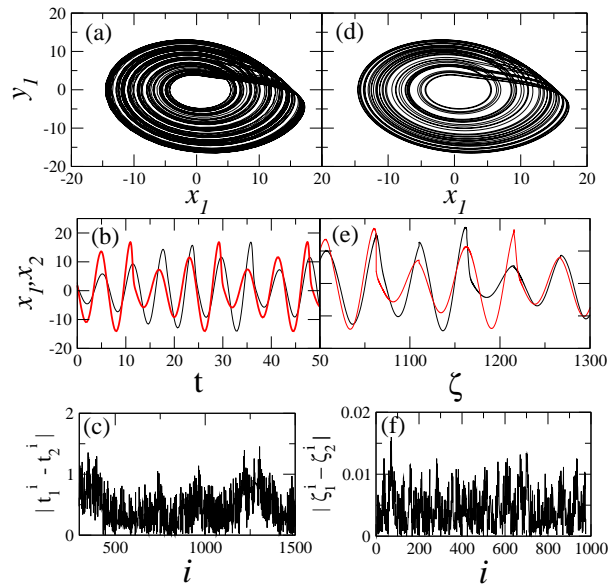,width=8.0cm}}}
\caption{The effect of time transformation in two synchronous R\"ossler oscillators.
  In (a) we show the attractor projection onto the $(x,y)$-plane, and
  in (b) the time series $x_1$ and $x_2$ vs. $t$.  In (c) we depict
  the time difference  $|t_1^i - t_2^i|$ for $i$th period between both oscillators. 
  One can see that even though the quantity
  $|t_1^i - t_2^i|$ is bounded, it has large fluctuations.  In
  (d-f) we proceed a time transformation $t \rightarrow \zeta$
  given by Eq. (\ref{TRros}). In (d) we show that attractor projection
  in the subspace $(x,y)$, and the time series $x_1$ and $x_2$ vs.
  $\zeta$ in (e), while the amount $|\zeta_1^i - \zeta_2^i|$ is shown
  in (f). One can see that after the time transformation the timing
  condition is drastically improved.}
\label{exemplo}
\end{figure}

We first illustrate our approach for the paradigmatic example of two
coupled R\"ossler oscillators: 
\begin{eqnarray}
\dot{x}_{1,2} & = & -\alpha_{1,2}y_{1,2}-z_{1,2}+\epsilon (x_{2,1}- x_{1,2}), \\
\dot{y}_{1,2} &=& \alpha_{1,2}x_{1,2}+0.15y_{1,2}, \\
\dot{z}_{1,2} &=& 0.2+z_{1,2}(x_{1,2}-10)
\end{eqnarray}
 with $\alpha_{1}=1$, and
$\alpha_{2}=\alpha_{1}+\Delta \alpha_2$. We shall denote ${\bf x}_{j}
= (x_j,y_j,z_j)$, where $j=1,2$, and ${\bf x} = ({\bf x}_1,{\bf x}_2,
\dot{\bf x}_1, \dot{\bf x}_2)$. Since for these oscillators the trajectory revolves 
around one specific point [Fig. (\ref{exemplo})(a)], we can simply define a phase by $\tan \phi_j =
y_j/x_j$, which yields \cite{tiagoPLA}
\begin{equation}
\phi_j({\bf x},t) = \int_0^t (\dot{y}_j x_j
- \dot{x}_j y_j)/(x^2_j + y^2_j) dt.
\end{equation}
 Furthermore, let us denote the time at which the oscillator ${\bf x}_j$
 completes its $i$th cycle by $t_{j}^i$. That is, the times at which the phase
 is increased by $2 \pi$ (see \ref{A} for more details). We can show that
 there is PS if, and only if, we have
\noindent
\begin{equation}
| t_{1}^i  - t_{2}^i | \le \kappa,
\label{time_diff}
\end{equation}
\noindent
where $\kappa$ is the minimum finite number that bounds the inequality. For more details
concerning this equivalence see \ref{A}. The value of
$\kappa$ shows how well paced both oscillators are. The smaller the
value of $\kappa$ the better the timing between $ {\bf x}_1$ and
${\bf x}_2$.

\noindent

For $\epsilon = 0.0015$ and $\Delta \alpha_2 = 0.001$, the two
oscillators are in PS, which means that the phase difference $\Delta
\phi = \phi_1(t) - \phi_2(t)$ is bounded for all times. Consequently,
Eq. (\ref{time_diff}) holds [Fig. \ref{exemplo}(b,c)].  In the PS
regime the oscillators have the same mean frequency, namely 
$$
\langle \dot{\phi}_1 \rangle_t = \langle \dot{\phi}_2 \rangle_t \approx 1.035,
$$
where $\langle \cdot \rangle_t$ is the time average with
respect to $t$.  The average period is given by 
$$\langle T_{j}
\rangle_t = 2\pi / \langle \dot{\phi}_j \rangle_t \approx 6.067.
$$

We have that $max|t_{1}^i - t_{2}^i|$ corresponds approximately to 
$\langle T_{j} \rangle_t /4$ [Fig.  \ref{exemplo}(c)], which can
be rather problematic for a reliable communication system based on
chaos synchronization, since the two oscillators do not reach the Poincar\'e 
section with a precise timing. See Ref. \cite{Lai} for a detailed discussion.

The main question is then what could happen when we transform the time. 
Could we produce an effective improvement of the timing? If we linearly 
scale the time introducing $\zeta = \alpha t$, then the average period transforms as 
$$
\langle T_j \rangle_{\zeta} = \frac{\langle T_j \rangle_t}{\alpha},
$$
while the timing 
$$
|\zeta_1^i - \zeta_2^i| = \frac{|t_1^i - t_2^i|}{\alpha}.
$$
Thus, there is no effective improvement of the timing with respect to the
average period, since
$$
\frac{|\zeta_1^i - \zeta_2^i|}{\langle T_j \rangle_{\zeta}} = \frac{|t_1^i - t_2^i|}{\langle T_j \rangle_{t}}.
$$

The situation can be altered to improve the timing condition in Eq.
(\ref{time_diff}) by using a nonlinear time transformation, namely $t
\rightarrow \zeta$ of the form:
\begin{equation}
 d\zeta = \lambda({\bf x},t) dt.
\label{trans}
\end{equation}
\noindent
Such a transformation may distort directly the synchronization
phenomenon acting on the times $t_{j}^i$. To improve the timing
between the oscillators given $\gamma \gg 1$ and $\sigma < 1$, we perform the time
transformation:
\begin{eqnarray}
\lambda({\bf x},t) = \left\{
\begin{array}{ccc}
\gamma, & \mbox{if} & x_{1,2}>0 \mbox{ and }x_{2,1}<0\mbox{ and }\dot{y}_1>0 \\
\sigma, & \mbox{otherwise}
\end{array}
\right.
\label{TRros}
\end{eqnarray}
which shrinks the time between $t_1^i$ and $t_2^{i}$ enhancing a more
accurate pacing between the oscillators.  $\gamma$ may be chosen
according to the pacing condition desired. For our purposes we
fix $\gamma =100$. We can use the parameter $\sigma$ to control the average period. In the following we fix 
$\sigma=0.11$.  The new time is given by $\zeta_j =
\int_{0}^{t_j} \lambda({\bf x},t) dt$.  The equation of motion now takes the form
\begin{eqnarray}
{\dot{x}}_{1,2} & = & \lambda^{-1}({\bf x},t)[ -\alpha_{1,2}y_{1,2}-z_{1,2}+\epsilon (x_{2,1}- x_{1,2}) ] , \\
{\dot{y}}_{1,2} &=& \lambda^{-1}({\bf x},t) [ \alpha_{1,2}x_{1,2}+0.15y_{1,2} ], \\
{\dot{z}}_{1,2} &=& \lambda^{-1}({\bf x},t) [0.2+z_{1,2}(x_{1,2}-10) ]
\end{eqnarray}

The time transformation causes no changes in the state space, compare
Figs. \ref{exemplo} (a) and (b).  However, the time series of $x_1
\times t$ and $x_1 \times \zeta$ are drastically modified [Fig.
\ref{exemplo}(b,e)].  Although, the time transformation is not able to
interfere with the PS phenomenon [Fig.  \ref{exemplo}(f)], it changes
the frequency of the oscillators $$ \langle \dot{\phi}_1
\rangle_{\zeta} = \langle \dot{\phi}_2 \rangle_{\zeta} \approx
0.998,$$ which implies that $\langle T_{1,2} \rangle_{\zeta} \approx
6.296$.  On the other hand, now $max|\zeta_{1}^i - \zeta_{2}^i|
\approx \langle T_{j} \rangle_{\zeta} / 420$. Remembering that
$max|t_{1}^i - t_{2}^i| \approx 2\langle T_{j} \rangle_{t} /3$, we
conclude that this time transformation yields an improvement of a
factor of $280$ for the timing.  Of course, Eq.  (\ref{TRros}) can be
altered to have an even better timing.  These ideas can also be
applied to a network.  Whenever there is a cluster of oscillators in
PS within the network, one can transform the time by Eq. (\ref{trans})
suitably choosing $\lambda({\bf x},t)$ to have a precise timing among
all oscillators of the PS cluster. 

\section{Phase Diffusion and Coherence}

Time transformation cannot destroy the synchronization.  On the
other hand, it does alter important characteristics of the dynamics,
for example the coherence of the oscillators and the phase diffusion.
By a time transformation we can transform an oscillator that
originally is endowed with phase diffusion into an oscillator with an arbitrarily small 
phase diffusion.  An interesting point is that in data analysis the
phase diffusion plays a role in order to detect PS
\cite{Lai2006}. The idea is that one can detect PS by variations in
the phase diffusion.

In general the phase depends on the amplitude of the oscillator and
the frequency can be written as: $\dot{\phi}(t) = \omega + \xi({\bf
  x},t),$ where $\omega$ is the average frequency of the oscillator and 
$\xi({\bf x},t)$, in many cases, acts as an effective noise due to the
chaotic nature of the oscillator \cite{livro}. Therefore, the phase dynamics is
generally diffusive, which means that for large time intervals one
expects $\langle [ \phi(t) - \omega t ]^2 \rangle_{\mu} \approx \Gamma
t$, where $\langle \cdot \rangle_{\mu}$ denotes the ensemble average,
and $\Gamma$ the diffusion constant.  Having the time of the $i$th
cycle of the oscillator ${\bf x}_j$, we can write $t_{j}^i = i \langle
T_{j} \rangle + \nu^i_{j}$, where $\langle T_{j} \rangle$ is the
average period. By calculating the phase diffusion, we have 
$$
\langle [
\phi_{j}( t_{j}^i ) - \omega_{j}( i \langle T_{j} \rangle + \nu^i_{j})
]^2 \rangle_{\mu} = \omega_{j}^2 \langle [ \nu^i_{j} ]^2 \rangle_{\mu}
\approx \Gamma_{j} t. $$ 
 Hence, $t_{\ell}^i- i \langle T_{\ell}\rangle$
gives the phase diffusion properties.

Let us analyze the distortions in the phase diffusion by a time
transformation and its effect on PS.  Supposing that the
oscillators ${\bf x}_1$ and ${\bf x}_2$ are not in PS, we write $
t_{1}^i - t_{2}^i = \alpha \times i + \xi^i$, where $\alpha, \xi^i
\in \mathbb{R}$ are chosen to hold the equality.  By performing a time
coordinate change we endow the oscillator ${\bf x}_1$ with zero phase
diffusion.  This means that we have a new time $\zeta$ with 
\begin{equation}
\Delta \zeta_1^i = \zeta_{1}^i - \zeta_1^{i-1} = 1,
\end{equation}
i.e. $\Delta \zeta_1^i =
\Delta t_{1}^i/ \Delta t_{1}^i$, where $\Delta t_{1}^i = {t}_{1}^i -
{t}_{1}^{i-1}$.  The new time coordinate is given by 
\begin{eqnarray}
\zeta_1^i &=&\sum_{n=0}^i \Delta t_1^n / \Delta t_1^n \nonumber \\
\zeta_2^i &=& \sum_{n=0}^i \Delta t_2^n / \Delta t_1^n. \nonumber
\end{eqnarray}

We have 
\begin{equation}
|\zeta_{1}^i -
\zeta_{2}^i|= | \sum_i (\Delta t_{1}^i - \Delta t_{2}^{i})/\Delta
t_{1}^{i}|. 
\end{equation}

 Next, consider the maximum $\Delta t_{1}^{i}$, namely
$$
max_i \Delta t_{1}^{i} = \gamma^{-1}.
$$ 
Thus, we have 

\begin{equation}
| \sum_i (\Delta t_{2}^i - \Delta t_1^i) / \Delta t_1^i | \ge \gamma | t_{2}^i
- t_{1}^i|
\end{equation}
which can be written as: 
\begin{equation}
| \zeta_{1}^i - \zeta_{2}^i| \ge \gamma (\alpha \times i + \xi^i ).
\end{equation}
 Therefore, as the number of
periods tends to infinity, the time event
difference $|\zeta_{1}^i - \zeta_{2}^i |$ diverges. Thus, enhancing
coherence in the oscillator does not introduce PS.

\section{Breaking down the Hypotheses on $\lambda$}

By violating the conditions $(ii)$ and $(iii)$, which guarantees the
boundedness of $\lambda({\bf x},t)$, PS can be introduced.
Considering our former case where 
\begin{equation}
 t_{1}^i - t_{2}^i = \alpha \times i + \xi^i, 
 \end{equation}
 we could transform the time by 
 \begin{equation}
 \lambda({\bf
  x},t)= \frac{1}{i} \, \, \mbox{if} \,\, t_1^i < t \le t_1^{i+1}.
 \end{equation}
  
The timing condition is
given by 
\begin{equation}
|\zeta_{1}^i - \zeta_{2}^i| = |\int_{t_1^i}^{t_1^i - \alpha
  i - \xi^i}\lambda({\bf x},t)| \le | \alpha i / i| + |\xi^i / i|.
\end{equation}  
 
Thus, 
\begin{equation}
\lim_{i \rightarrow \infty}| \zeta_{1}^i - \zeta_{2}^i| \le
\alpha,
\label{deltazeta}
\end{equation}
the time difference is bounded by $\alpha$. Hence, this time
transformation allows us to introduce PS between ${\bf x}_1$ and
${\bf x}_2$. This does not contradict our results, because this time
transformation is not bounded, violating the assumptions $(ii)$ and
$(iii)$.  When, one uses a non-bounded transformation $\lambda({\bf
  x},t)$, like the latter one, the time is shrunk and becomes
meaningless; there is no long term behavior with respect to $\zeta$,
since $\lim_{i \rightarrow \infty} \zeta_{\ell}^i$ is still bounded. 
The function $\lambda({\bf x},t)$ can be made smooth without changing
Eq. (\ref{deltazeta}). Here we have considered $\lambda({\bf x},t)$ a step 
function without lost of generality. 

\section{Feigning Phase Synchronization}

\begin{figure}[h]
  \centerline{\hbox{\psfig{file=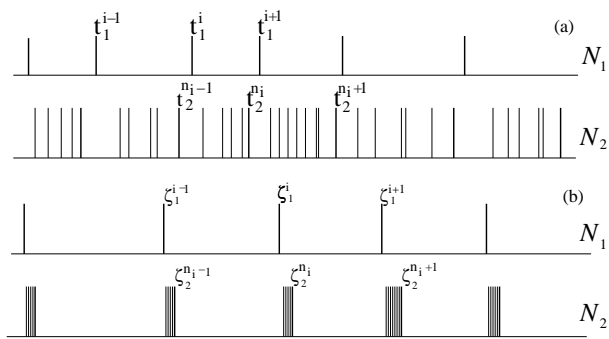,width=8.0cm}}}
\caption{The effect of time transformation in two asynchronous neurons.
  In (a) the spikes in both neurons are governed by stochastic
  processes.  They present no synchronization.  From (a) to
  (b) we perform the transformation given by Eq. (\ref{exe}), which
  shrinks the time between $t_1^i$ and $t_2^{n_i}$ and stretches the
  time between $t_2^{n_i}$ and $t_1^{i+1}$.  As a consequence, it
  seems that phase synchronization between the spikes in $\mathcal{N}_1$ and
  the bursts of $\mathcal{N}_2$ is enhanced.}
\label{exemplo2}
\end{figure}

PS is invariant whenever $\lambda({\bf x},t)$ fulfills the conditions
$(i-iii)$. However, it is important to mention that the phase
definition must be defined  consistently. This means that one must have the
same phase definition before and after the time transformation,
otherwise one could predict that PS is not invariant, due to the
changing of the phase definition. Let us consider two spiking neurons $\mathcal{N}_1$ and
$\mathcal{N}_2$. We assume that the spike times $t_1^i$ and $t_2^i$
are independent, with neuron $\mathcal{N}_1$ having a
higher frequency, Fig.  \ref{exemplo2}(a), in such a way that there is
no $n:m$ PS between $\mathcal{N}_1$ and $\mathcal{N}_2$.  Let
$t_2^{n_i}$ be the spike time of $\mathcal{N}_2$ that precedes the
$i+1$th spike of $\mathcal{N}_1$, Fig.  \ref{exemplo2}(a). Then, given
$\sigma_i \ll \gamma_i$, we perform the following transformation:
\noindent
\begin{eqnarray}
\lambda(t) = \left\{
\begin{array}{ccc}
\sigma_i , & \mbox{if} & t_1^i < t \le t_2^{n_i} , \\
\gamma_i, & \mbox{if} &   t_2^{n_i} < t \le t_1^{i+1}. \\
\end{array}
\right.
\label{exe}
\end{eqnarray}
\noindent
This shrinks the time between $t_1^i$ and $t_2^{n_i}$ and stretches
between $t_2^{n_i}$ and $t_1^{i+1}$ creating bursts in
$\mathcal{N}_2$.  In the rescaled time $\zeta$, the bursts of
$\mathcal{N}_2$ is synchronized with the spikes of $\mathcal{N}_1$.
However, there is no synchronization between the spikes, since
$\lim_{i \rightarrow \infty}|\zeta_1^i - \zeta_2^i| \rightarrow
\infty$. On the other hand, after the time coordinate change, it is
very tempting to introduce a phase for the bursts that increases
$2\pi$ between two successive bursts of $\mathcal{N}_2$.  Changing the
phase definition, the phase difference between $\mathcal{N}_1$ and
$\mathcal{N}_2$ becomes bounded. Therefore, it seems 
possible to introduce PS between two asynchronous neurons. However,
this is a fake PS once that the phase definition is changed.

\noindent

\section{Network Information Transmission}

Let us analyze the effect of time transformations in the information transmission in
networks.   For every pair of oscillators ${\bf x}_j$ and ${\bf x}_k$ we can define a
coordinate transformation 
\begin{eqnarray}
{\bf x}^{\parallel}_{jk}={ \bf x}_j+{\bf x}_k \label{coord_parallel} \\ 
{\bf x}^{\perp}_{jk}={\bf x}_j - {\bf x}_k, \label{coord_perp}
\end{eqnarray}
 that produces two positive conditional
Lyapunov exponents (in units of bits/unit time) $\sigma^{\parallel}(t)$ and
$\sigma^{\perp}(t)$. The mutual information rate (MIR)  is
bounded from above \cite{Murilo-Canal}
\begin{equation}
I_C(t)  \le \sigma^{\parallel}(t)_-\sigma^{\perp}(t)
\end{equation}

The main goal is to know how the mutual information rate behaves as
we implement a time transformation.  By choosing a proper nonlinear 
$\lambda({\bf x},t)$ in Eq. (\ref{trans}) we can introduce different
time scales in the oscillators time series as well as endow the time transformation with as much information
about the dynamics as we want. The main question is whether, under such 
nonlinear $\lambda({\bf x},t)$, the information contained in 
$\lambda({\bf x},t)$ could be transmitted to the oscillators.

To answer this question we need to uncover the general transformation law for $I_C$.
After some manipulations we can uncover 
the transformation law of $I_C(t)$:
\begin{equation}
I_C(\zeta) \leq \frac{I_C(t)}{\langle \lambda \rangle_t},
\label{I_C_trans}
\end{equation}
where, again $\langle \cdot \rangle_t$ stands for  the time average. For details see \ref{B}.

Equation (\ref{I_C_trans}) shows an invariant character of $I_C$.  {\it if
$I_C(t)=0$, what happens in the absence of synchronization (correlation)
between oscillators, no time transformation that respects conditions $(i-iii)$
can raise $I_C(t)$ from zero.}  Hence, no matter how much information is
contained in $\lambda({\bf x},t)$, if there is no synchronization this
information cannot be used.  If, on the other hand, $I_C(t)$ is positive, then
$I_C(\zeta)$ can be made arbitrarily large.

To illustrate our findings, we consider a network
of  four identical  Hindmarsh-Rose chaotic neurons electrically coupled in an all-to-all topology,
\begin{eqnarray}
\dot{x}_j &=& y_j + 3x_j^2 - x_j^3-z_j + I_j + \sum_k C_{jk}(x_k-x_j), \\
\dot{y}_j &=& 1-5x_j^2-y_j, \\
\dot{z}_j  &=& -rz_j + 4r(x_j+1.6),
\end{eqnarray}
 where $C_{j k}$ 
stands for the coupling matrix. We use
$r$=0.005, $I_i=3.2$, and random initial conditions. 

\begin{figure}[h]
  \centerline{\hbox{\psfig{file=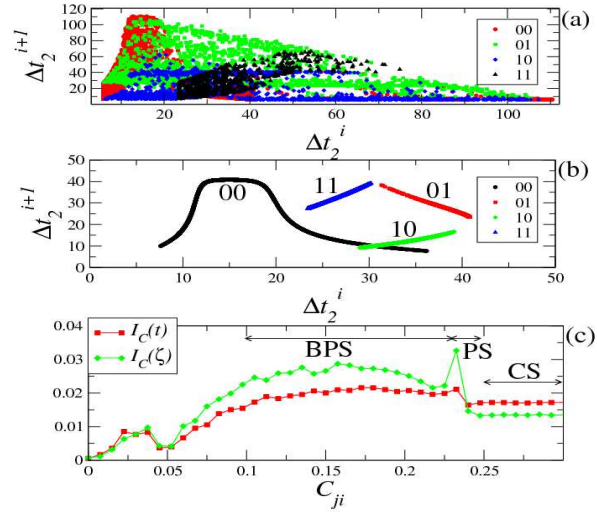,width=8.0cm,height=7.0cm}}}
\caption{[Color online] Encoded time intervals between two spikes in neuron 
${\bf x}_2$ for $C_{jk}$=0 (a), and for $C_{jk}$=0.3 (b). In (c), the MIR 
between neurons ${\bf x}_1$ and ${\bf x}_2$, for the
time-$t$ (filled squares), and for the time-$\zeta$ (filles diamonds). 
BPS (burst phase synchronization) is found  for $C_{jk}$=[0.1,0.23].
In this regime only the burst are phase synchronized. PS is found for
$C_{jk}$=[0.23,0.25], and CS (complete synchronization) is found 
for $C_{jk}$=[0.25,0.3]. For $C_{jk} \in $ [0.05,0.23] $\lambda({\bf x},t)$ is smaller than 1, which 
provides an increasing in $I_C$ up to $60\%$. For CS $\langle \lambda \rangle _t $ = 1.29 
providing a decreasing in $I_C$.}
\label{TR_fig1}
\end{figure}
\noindent

We define the following time transformation
\begin{eqnarray}
\lambda({\bf x}_1,t) = \left\{
\begin{array}{ccc}
\alpha, & \mbox{if} & x_{1}=0 \mbox{ and }y_{1}>-4.6 ,\\
\beta, & \mbox{if} & x_{1}=0 \mbox{ and }y_{1} \leq -4.6. \\
\end{array}
\right.
\label{TRhr}
\end{eqnarray}
\noindent

This shrinks the time between spikes  when the $y_1 > -4.6$ and stretches
when $y_1 > -4.6$ creating a frequency modulation between the spikes, which depends 
on the trajectory position. Hence, the 
transformation carries information about ${\bf x}_1$.
In our analysis we keep fix $\alpha = 0.5$ and $\beta = 2$. $t_j^i$ denotes the time of 
 the $i$th crossing of the trajectory of ${\bf x}_j$ with the
section $x_j=0$ (an spike event). The time interval between two crossings is
$\Delta t_j^i=t_j^{i+1}-t_j^{i}$.
 
We introduce a symbolic dynamics which exhibits rather easily the 
results for the distinct synchronization regimes.  We can encode the 
binary information about the transformation $\lambda({\bf x}_1,t)$ 
by setting $\alpha$ to the symbol "0" and $\beta$ to "1". Hence,   
we have for two consecutive $\lambda = \alpha $:  '00'; one 
$\lambda = \alpha$ followed by $\lambda =\beta$: '01'; one 
$\lambda =\beta$ followed by $\lambda = \alpha$:  '10';  and finally 
two consecutive $\lambda = \beta$: '11'.  Whenever the time transformation
is able to transmit the information about the symbols, we can access the
information about ${\bf x}_1$ in the spike time intervals of the other neurons.

Figures \ref{TR_fig1}(a-b) show return maps $\Delta t_2^i$ vs. $\Delta
t_2^{i+1}$ of the neuron ${ \bf x}_2$. We split this map into four return maps, 
depending on the value of the transformation
$\lambda({\bf x}_1,t)$. That is, distinguished by the different symbols '00','01','10','11'.
The information about the values of $\lambda({\bf x}_1,t)$ should be considered to be
unknown, but here we make use of it to illustrate our ideas.

By measuring $\Delta t_2^i$ we should be able to infer 
the time interval $\Delta t_1^i$, if the time transformation can transmit information.
Figure \ref{TR_fig1}(a) shows that return maps $\Delta t_2^i \, vs. \, \Delta t_2^{i+1}$ for the different values of 
$\lambda({\bf x}_1,t)$ superimpose, and as a
consequence it is impossible to discern whether the region that encodes for 00
is mapped to either 01 or 00, and so on. That leads to a complete uncertainty about
$\lambda({\bf x}_1,t)$ by measuring $\Delta t_2^i$. Therefore, there is
no exchange of information between ${\bf x}_1$ and ${\bf x}_2$.

The time scale of ${\bf x}_2$ is being rescaled according to a function that
contains information about the position of ${\bf x}_2$. From the way the
function $\lambda$ is constructed, whenever $y_1>-4.6$ and $x_1=0$, the
oscillation frequency of the oscillator ${\bf x}_2$ in the time-$\zeta$ frame
is increased. Whenever $y_1 \le -4.6$ and $x_1=0$, the oscillation frequency
of the oscillator ${\bf x}_2$ in the time-$\zeta$ frame is decreased.  So, the
oscillation frequency of ${\bf x}_2$ is being modulated. Frequency modulation
(FM) is a typical procedure to transmit information, a protocol in which the
information signal is carried by the frequency of a wave. It would be
natural to imagine that by modulating the oscillator ${\bf x}_2$ using a time
transformation based on the position of ${\bf x}_1$ one could realize at least
partially information about ${\bf x}_1$ by making measures in ${\bf
x}_2$. However, surprisingly, that is not the case in dynamical networks.
Therefore, if elements in a dynamical network do not exchange information
among themselves, there is no time transformation that can change this scenario.

When the neurons are completely synchronized (for $C_{jk}$=0.3), we see in
Fig. \ref{TR_fig1}(b) that except for one point, the return maps $\Delta t_2^i
\, vs. \, \Delta t_2^{i+1}$ for different values of $\lambda({\bf x}_1,t)$ are
disjoint, which means that by measuring $\Delta t_2^i$ we have complete
knowledge about the trajectory of the neuron ${\bf x}_1$.

\subsection{Effect of $\lambda({\bf x}_1,t)$ on $I_C(t)$}

We keep fix $\lambda({\bf x}_1,t)$ and vary the coupling strength
$C_{jk}$. Equation (\ref{I_C_trans}) states that whenever $\langle
\lambda({\bf x}_1,t) \rangle_t < 1$ the time transformation increases the MIR.
In Fig. \ref{TR_fig1}(c) we show the MIR between $\Delta t_1^{i}$ and $\Delta
t_2^{i}$ using the Shannon mutual information \cite{encoding}, for the two
time frames. $I_C(t)$ denotes the MIR in the time-$t$ frame and $I_C(\zeta)$
the MIR in the time-$\zeta$ frame. For $C_{ij} \in [0.5, 0.23]$ $\langle
\lambda({\bf x}_1,t)
\rangle_t < 1$ which provides an effective increasing in the MIR. 

In Eq. (\ref{TRhr}), $\lambda$ is defined to contain information about
${\bf x}_1$. However, $\lambda$ could be defined to contain
information about an arbitrary information signal to be
transmitted. In such a case, each disjoint region [as the ones shown
in Fig. \ref{TR_fig1}(b)] would encode information about this signal,
which can be retrieved somewhere else in the network.

$\lambda({\bf x},t)$ can be
constructed using information about some particular node of the
network, a group of nodes. Whenever the oscillators are phase 
synchronized, we can improve  the mutual information rate 
by using  $\lambda({\bf x},t)$ that contains information about the 
dynamics of the phase synchronized oscillators.

\section{Conclusions}

In summary, we have shown that for general dynamical oscillators it is
neither possible to introduce nor to destroy PS by a time
transformation. Furthermore, we have discussed possible application of
these ideas to relevant technological problems such as nonlinear
digital communication \cite{Lai}. Moreover, we have illustrated
these results for nonsynchronized oscillators, showing that the
enhancement of zero phase diffusion does not enhance PS. We have also
discussed that breaking the boundedness condition imposed on $\lambda$
PS can be enhanced. However such a transformation is physically
meaningless.  Finally, we have shown that the time transformation
can introduce the presence of distinct time scales, which can feign
PS. Our findings might be relevant to several areas of natural science
for the study of synchronization where the exact time the phenomenon 
took place is unknown and only a proxy for the time can
be derived from the measurements. Examples can be
found in geophysics when sediment cores are studied. 
Such situations may arise in dendrochronology, ice cores and  three rings.

{\bf Acknowledgment} We would like to thank M. B. Reyes, M.  Romano 
for a critical reading of the first version of the manuscript. This work was financially
supported by Helmholtz Center for Mind and Brain Dynamics,
the SPP 1114 of the "Deutsche Forschungsgesellschaft", and FAPESP.

\appendix

\section{PS invariance under time transformations} \label{A}

We consider two general oscillators $\dot{{\bf x}}_j = {\bf F}_j({\bf
  x}_j)$, where ${\bf x}_j \in \mathbb{R}^{n}$ and ${\bf
  F}_j:\mathbb{R}^{n} \rightarrow \mathbb{R}^{n}$, and analyze a
general coupling scheme:
\noindent
\begin{equation}
\dot{{\bf x}}_{1,2} = {\bf F}_{1,2}({\bf x}_{1,2}) + {\bf C}_{1,2}({\bf x},t)[{\bf
  H}_{2,1}({\bf x}) - {\bf H}_{1,2}({\bf x})],
\label{ds}
\end{equation}
\noindent
where ${\bf H}_j({\bf x})$ is the coupling vector function, and ${\bf
  C}_{j}({\bf x},t)$ is the coupling matrix. Note that this scheme
also takes unidirectional couplings (master-slave configuration) into
account. We suppose that each ${\bf x}_j$ has a stable attractor and a
frequency $ \dot{\phi}_j = \Omega_j({\bf x},t),$ where $\Omega_j({\bf
  x},t)$ is a continuous function (or Riemann integrable).
Furthermore, we assume that there is a number $M$ such that
$\Omega_j({\bf x},t) \le M$. From now on, slightly abusing the
notation we shall omit the dependence of the functions on the
coordinates and on time, whenever there are no problems with the
notation.  Given a finite real number $c$, the condition for PS between ${\bf
  x}_1$ and ${\bf x}_2$ can then be written as:
\begin{equation}
|\phi_1(t) - \phi_2(t)| < c.
\end{equation}
First, we formalize the relation between PS and the timing condition
given by Eq. (\ref{time_diff}). We have: 
$$
| \phi_1(t) -
  \phi_2(t) | = | \int_0^{t_1^i} \Omega_1 dt - \int_0^{t_2^i} \Omega_2
  dt - \int_{t_2^i}^{t_1^i} \Omega_2 dt + \beta^i(t) |$$

where 
$$\beta^i(t) = \int_{t_1^i}^t \Omega_1 dt - \int_{t_1^i}^t \Omega_2 dt.$$

Next, since $\phi_{j}(t_j^i)$ is equal to $i \times 2\pi$ \cite{phi0}, it yields:
\begin{equation}
| \phi_1(t) - \phi_2(t) | \le  M | t_1^i - t_2^i | + max_i | \beta^i(t) |. \\
\end{equation}
\noindent

The term $max_i | \beta^i |$ is always bounded.
By hypothesis we have 
$$
|t_{j}^i - t_{j}^{i-1}| \le  \Lambda
$$ 

we have 

\begin{eqnarray}
| \beta^i | &=& \left|  \int_{t_1^i}^t \Omega_1 dt - \int_{t_1^i}^t \Omega_2 dt \right| \\
& \le & \left|  \int_{t_1^i}^t \Omega_1 dt \right| +  \left| \int_{t_1^i}^t \Omega_2 dt \right| 
\end{eqnarray}

But now remembering that $max_{j,t}\{  \Omega_j(t) \} = M$, then

\begin{eqnarray}
| \beta^i | \le 2 M \Lambda
\end{eqnarray}

Therefore, a bounded time event difference $| t_1^i - t_2^i |$ implies
the boundedness of the phase difference.  A similar argument shows
that the boundedness of the phase difference implies the boundedness
of the time event difference.  Therefore, Eq.  (\ref{time_diff}) is
equivalent to PS.

We analyze the effect of time transformation PS. We assume
$\lambda({\bf x},t)$ to be $(i)$ {\it at least Riemann integrable}
$(ii)$ {\it finite}, and $(iii)$ {\it bounded away from zero}.  The
two latter conditions are equivalent to the existence of two numbers
$\delta^{-1},\eta \in\mathbb{R}_{+}$ such that $\delta^{-1} \le
\lambda({\bf x},t) \le \eta$.  Under the assumptions $(i-iii)$ we can
demonstrate that PS is invariant under time transformations. First, we
show that 
$$| t^i_1 - t_2^i | \le \kappa \Rightarrow | \zeta^i_1 - \zeta_2^i | \le \tilde \kappa
$$

Noting that 
  $\zeta_1^i = \int_{0}^{t_1^i} \lambda({\bf x},t)dt$ and $\zeta_2^i = \int_{0}^{t_2^i}\lambda({\bf x},t)dt $, we find

$$ 
| \zeta_1^i - \zeta_2^i | = \left| \int_{0}^{t_1^i} \lambda({\bf x},t)dt - \int_{0}^{t_2^i}\lambda({\bf x},t)dt \right|.
$$

This may be
written as $| \zeta^i_1 - \zeta_2^i |= | \int_{t_2^i}^{t_1^i}
  \lambda({\bf x},t)dt |$.  However, since $\lambda({\bf
    x},t) \le \eta$ we have 
$$| \zeta_1^i - \zeta_2^i | \le \eta |
t_1^i - t_2^i |.
$$ 
Thus, the boundedness of $| t_1^i - t_2^i |$ implies
the boundedness of $| \zeta_1^i - \zeta_2^i |$. Now, we show that
$$
| \zeta^i_1 - \zeta_2^i | \le \tilde \kappa \Rightarrow | t^i_1 - t_2^i | \le \kappa.
$$  

We have
$$
| t_1^i - t_2^i | =  \left| \int_{0}^{\zeta_1^i} \lambda^{-1}({\bf x},t) dt - \int_{0}^{
    \zeta_2^i} \lambda^{-1}({\bf x},t) dt \right|,
$$ 

which equals $|
  \int_{\zeta_2^i}^{\zeta_1^i} \lambda({\bf x},t)^{-1} dt|$.  As
$\delta^{-1} < \lambda({\bf x},t)$, we get 

$$
| t_j^1 - t_2^i | \le \delta | \zeta_1^i - \zeta_2^i |
$$ 

Therefore, we conclude that
there is PS in the "new" time frame $\zeta$ if and only if there is PS
in the original time $t$.

The results stated in this section are general and do not depend on the
attractor topology or coherent properties, as long as a phase can be
introduced. Note that we do not have to know the phase equation, but
only assume that it exists.

The onset of phase synchronization, and even the phase equation, depends on
the attractorsÕ topology and coherence \cite{livro}. If the attractor has a
simple topology, that is, it has proper rotation, then, the onset of phase
synchronization is given by a transition of the zero Lyapunov exponent to
negative values. For such a case, the results of Ref. \cite{Adilson},
concerning the invariance of the sign of the Lyapunov exponents, can be used
to state the invariance of PS under time transformations.

\section{Transformation Law for  Mutual Information Rate} \label{B}

Representing one node dynamics of the network by 
\begin{equation}
\frac{d {\bf x}}{dt} = {\bf F}({\bf x}),
\label{network_louca}
\end{equation}
the Lyapunov exponents of an invariant set of the phase space
are defined as
\begin{equation}
h_t^i = \lim_{t\rightarrow \infty} \ln \frac{| {\bf y}_t^i |}{| {\bf y}_{t0}^i |},
\end{equation}
with
\begin{equation}
\dot{{\bf y}}_t^i = {\bf D} {\bf F} {\bf y}_t^i.
\end{equation}

${\bf x}(0)$ is a typical initial condition and ${\bf y}_{t0}^i = {\bf
y}_{t}^i(0)\sigma^{\perp}_{t}$ are the tangent vector at ${\bf x}(0)$. In Ref. \cite{Adilson} it
is shown the transformation law for the Lyapunov spectrum. Under a time
transformation $\lambda$ fulfilling the hypotheses $(i-iii)$ we have
\begin{equation}
h_{\zeta}^i = \frac{h_t^i}{ \langle \lambda \rangle_t }.
\end{equation}

Here, we show that the conditional Lyapunov exponents
$\sigma_{\parallel}$ and $\sigma_{\perp}$ follow the same
transformation law, since the conditional Lyapunov exponents are the
Lyapunov exponents of the network considering that all the initial
conditions are equal, and therefore, the same results from
\cite{Adilson} apply.

Expanding Eq. (\ref{network_louca}) linearly around the synchronous
state $s$, and using the parallel coordinate defined in
(\ref{coord_parallel}), we arrive that

\begin{eqnarray}
\dot{{\bf x}}^{\parallel} &=& 2 [{\bf F}({\bf s}) - 
{\bf D}{\bf F}({\bf s}) {\bf s} ] + {\bf D}{\bf F}({\bf s}) {\bf
x}^{\parallel} \\ &=& {\bf G}({\bf s}, {\bf x}^{\parallel}).
\label{e1}
\end{eqnarray} 
\noindent
 
Proceeding in the same way for the perpendicular coordinate in
(\ref{coord_perp}), we arrive that

\begin{eqnarray}
 \dot{{\bf x}}^{\perp} &=& {\bf D}{\bf F}({\bf s}) {\bf x}^{\perp} \\
 &=& {\bf M}({\bf s},{\bf x}^{\perp}).
\label{e2}
 \end{eqnarray}

Now by means of ${\bf G}$ and ${\bf M}$ one can obtain the variational
equations of Eqs. (\ref{e1}) and (\ref{e2}), which provide the way
small perturbations propagate along the parallel and perpendicular
directions. From them, we obtain the Lyapunov conditional exponent
$\sigma^{\parallel}_t$ along the parallel direction and the Lyapunov
conditional exponent $\sigma^{\perp}_t$ along the transversal
direction, respectively.

Then, by applying the results of Ref. \cite{Adilson}, we conclude that
\begin{eqnarray}
\sigma^{\parallel}_{\zeta} = \frac{\sigma^{\parallel}_{t}}{\langle
\lambda \rangle_t}, \label{Sigpa} \\ \sigma^{\perp}_{\zeta} =
\frac{\sigma^{\perp}_{t}}{\langle \lambda \rangle_t} .
\label{Sigpe}
\end{eqnarray}

While the parallel conditional exponents are the Lyapunov exponents of
the synchronization manifold, whose positive exponents measure the
rate of information produced by the nodes if they were completely
synchronous, the transversal conditional exponents are the Lyapunov
exponents along the directions transversal (orthogonal) to the
synchronization manifold, whose positive exponents measure the rate
of information that can be erroneously transmitted between nodes.  

For a matter of simplicity in the notation, we denote the sum of all
the positive parallel exponents by $\sigma^{\parallel}_t$ and the sum
of all positive transversal exponents by $\sigma^{\perp}_t$.
 
Then, by using the results of Ref. \cite{Murilo-Canal}, we can write

\begin{equation} 
I_C(t) \le \sigma^{\parallel}_{t} - \sigma^{\perp}_{t}, 
\label{lalala}
\end{equation} 
\noindent 
which in other words means that the mutual information rate, i.e. the
rate with which information is exchanged between two nodes of the
network, is given by the rate of information produced by the
synchronous trajectories minus the rate of information produced by the
desynchronous trajectories, the error in the transmission of
information.

To intuitively understand Eq. (\ref{lalala}), one can compare the
right hand side of it with the usual definition of mutual information
rate between a source of information, denoted by $S$, and the receiver
of information, denoted by $R$, given by $H(S)-H(S|R)$. The term
$H(S)$, which can be compared with $\sigma^{\parallel}_{t}$,
represents the rate with which information is produced in the source
and the term $H(S|R)$, which can be compared with
$\sigma^{\perp}_{t}$, represents the rate of uncertainty remaining
about the transmitted information after observing the received
information, i.e., the rate with which information is erroneously
transmitted.

Then, taking into account Eqs. (\ref{Sigpa}) and (\ref{Sigpe}) we have
\begin{equation}
I_C(\zeta) \le \frac{1}{\langle \lambda \rangle_t} ( \sigma^{\parallel}_{t} - \sigma^{\perp}_{t}),
\end{equation}
concluding
\begin{equation}
I_C(\zeta) \le \frac{I_C(t)}{\langle \lambda \rangle_t}.
\end{equation}


\begin{thebibliography}{99}
\bibitem{phi0} Note that this is also true for $t_j^1$. This means that $\phi_j(t) = \int_0^{t_j^1} = 2 \pi$.
Thus, in this construction, there is no contribution of the initial phase $\phi_j(0)$.
\end{thebibliography}
\end{document}